\documentclass[iop,ap]{emulateapj}

\usepackage{aas_macros}
\usepackage{apjfonts}
\usepackage{color}
\usepackage{amsmath}
\usepackage{amssymb}
\usepackage{multirow}
\usepackage[caption=false]{subfig}
\usepackage{enumitem}
\setlist{nolistsep}
\usepackage{graphicx}	
\usepackage{bm}		
\usepackage{apjfonts}
\usepackage[plainpages=false, colorlinks=true, anchorcolor=blue, linkcolor=blue, citecolor=blue, bookmarks=false]{hyperref}
\usepackage{comment}
\usepackage{url}
\usepackage[T1]{fontenc}
\usepackage{ae,aecompl}
\usepackage{newtxtext}

\newcommand\finetilde{{\raise.17ex\hbox{$\scriptstyle\sim$}}} 

\def\be{\begin{equation}}
\def\ee{\end{equation}}

\def\apjl{{\it Astrophys. J.} Lett~}

\begin{document}

\title[Are we there yet?]{Are we there yet? \\ Time to detection of nanohertz gravitational waves based on pulsar-timing array limits}

\author{
S.~R.~Taylor\altaffilmark{1,2}\altaffilmark{$\star$}, 
M.~Vallisneri\altaffilmark{1,2}, 
J.~A.~Ellis\altaffilmark{1,2}\altaffilmark{,$\dagger$}, 
C.~M.~F.~Mingarelli\altaffilmark{2, 3, 1},
T.~J.~W.~Lazio\altaffilmark{1,2}, \&
R.~van~Haasteren\altaffilmark{1,2}\altaffilmark{,$\dagger$}
}

\altaffiltext{$\star$}{\email[email: ]{Stephen.R.Taylor@jpl.nasa.gov}}
\altaffiltext{1}{Jet Propulsion Laboratory, California Institute of Technology, 4800 Oak Grove Drive, Pasadena, CA 91109, USA}
\altaffiltext{2}{TAPIR Group, MC 350-17, California Institute of Technology, Pasadena, California 91125, USA}
\altaffiltext{3}{Max Planck Institute for Radio Astronomy, Auf dem H\"{u}gel 69, D-53121 Bonn, Germany }
\altaffiltext{$\dagger$}{Einstein Fellow}


\begin{abstract}
Decade-long timing observations of arrays of millisecond pulsars have placed highly constraining upper limits on the amplitude of the nanohertz gravitational-wave stochastic signal from the mergers of supermassive black-hole binaries ($\sim 10^{-15}$ strain at $f = 1/\mathrm{yr}$).
These limits suggest that binary merger rates have been overestimated, or that environmental influences from nuclear gas or stars accelerate orbital decay, reducing the gravitational-wave signal at the lowest, most sensitive frequencies. 
This prompts the question whether nanohertz gravitational waves are likely to be detected in the near future. In this letter, we answer this question quantitatively using simple statistical estimates, deriving the range of true signal amplitudes that are compatible with current upper limits, and computing expected detection probabilities as a function of observation time.
We conclude that small arrays consisting of the pulsars with the least timing noise, which yield the tightest upper limits, have discouraging prospects of making a detection in the next two decades. 
By contrast, we find large arrays are crucial to detection because the quadrupolar spatial correlations induced by gravitational waves can be well sampled by many pulsar pairs.
Indeed, timing programs which monitor a large and expanding set of pulsars have an $\sim 80\%$ probability of detecting gravitational waves within the next ten years, under assumptions on merger rates and environmental influences ranging from optimistic to conservative. Even in the extreme case where $90\%$ of binaries stall before merger and environmental coupling effects diminish low-frequency gravitational-wave power, detection is delayed by at most a few years.
\end{abstract}



\maketitle

\section{Introduction}
\label{sec:intro}

For the last ten years, three international collaborations have been collecting precise timing observations of the most stable millisecond pulsars, with the aim of identifying the imprint of gravitational waves (GWs) in the nanohertz frequency band \citep{h13,kc13,ml13}. The best-motivated GW source for such \emph{pulsar-timing arrays} (PTAs) is the stochastic \emph{background} signal from the cosmological population of gravitationally bound supermassive black-hole--binaries (SMBHBs: \citealt{rm95,jb03,wl03}) that are expected to form at the centers of galaxies after these merge \citep{shm+04,svc08}. If the binary inspirals are driven purely by GW radiation reaction at orbital separations corresponding to the PTA frequency band, the resulting timing-residual signal has a power-law spectrum with a characteristic exponent,
\begin{equation}
S(f) = \frac{h_c(f)^2}{12\pi^2 f^3} = \frac{A_{\rm GW}^2}{12 \pi^2} \left(\frac{f}{\mathrm{yr}^{-1}} \right)^{-13/3} \, \mathrm{yr}^3 \, ,
\label{eq:powerlaw}
\end{equation}
where $h_c(f)$ is the characteristic strain spectrum of the GW background. Theoretical expectations for the GW amplitude $A_\mathrm{GW}$ center around a few $10^{-15}$ \citep{mop14,s13,rws+14}. Recent observational upper limits from the three international pulsar-timing consortia \citep{ltm+15,arz+15,s+15} imply $A_\mathrm{GW} \lesssim 10^{-15}$ at 95\% confidence. These limits appear in tension with theoretical expectations, so much so that it seems plausible that environmental effects in galactic centers either stall the formation of gravitationally bound SMBHBs, or accelerate their inspiral; in both cases, the effective $A_\mathrm{GW}$ is reduced at the frequencies where PTAs are most sensitive \citep{arz+15,s+15}.

Indeed, the PTA detection of the stochastic GW background (GWB) from SMBHBs, considered imminent only a few years ago \citep{sejr13,mop14}, now seems to be receding toward the future. Is this really the case? In this letter, we answer this question quantitatively. Namely, given the upper limit $A_\mathrm{ul} = 10^{-15}$ obtained by the Parkes PTA (PPTA; \citealt{s+15}), we ask \emph{when} we can expect to make a positive detection using different PTAs: the PPTA \citep{h13}, limited to the four low--timing-noise pulsars used for the upper limit; the North American Nanohertz Observatory for Gravitational-waves (NANOGrav; \citealt{ml13}); the European PTA (EPTA; \citealt{kc13}); the International PTA (IPTA; see, e.g., \citealt{haa10}); and the pulsar-timing project that will be supported by SKA1 \citep{2015aska.confE..37J}.

We adopt the frequentist formalism developed by \citet{rsg15}, and we characterize the PTAs simply by listing, for each pulsar, the duration of the observation and the levels of measurement and timing noise. 
While GW searches and upper limits with PTAs have recently been given a Bayesian treatment \citep{vlml09,vl13,l+13}, the frequentist approach based on \emph{optimal statistics} is both convenient and appropriate for the purpose of this paper, because it dispenses with the simulation of actual datasets, and because the quantification of detection probability is intrinsically a frequentist statement (see, e.g., \citealt{2012PhRvD..86h2001V}). Furthermore, experience shows that frequentist upper limits and detection prospects are rather close to Bayesian results (see, e.g., \citealt{arz+15}).

We proceed as follows: for a specified PTA and true GW background amplitude ($A_\mathrm{true}$), we compute the detection probability (DP) as a function of observation time, using the detection statistic described below. Detection is a probabilistic endeavor since the actual realization of measurement and timing noise may obscure or expose the underlying GW signal. 
Indeed, we do \emph{not} currently have access to $A_\mathrm{true}$, but only to the observed upper limit $A_\mathrm{ul}$.
Setting upper limits is also a probabilistic endeavor, because different realizations of noise lead to different $A_\mathrm{ul}$ given the same $A_\mathrm{true}$, so we use the upper-limit optimal statistic (also described below) to compute $p(A_\mathrm{ul}|A_\mathrm{true})$. By introducing an astrophysically motivated prior $p(A_\mathrm{true})$, we can then use Bayes' theorem to obtain $p(A_\mathrm{true}|A_\mathrm{ul}) \propto p(A_\mathrm{ul}|A_\mathrm{true}) p(A_\mathrm{true})$. Finally, we obtain the \emph{expected} detection probability $\mathrm{DP}(A_\mathrm{ul})$ as $\int \mathrm{DP}(A_\mathrm{true}) p(A_\mathrm{true}|A_\mathrm{ul}) \, \mathrm{d}A_\mathrm{true}$.\footnote{By assuming a prior, we are introducing a Bayesian element in a frequentist scheme, but this is necessary because we wish to make a statement about a \emph{range} of astrophysical possibilities, which are all compatible to varying degrees with the observed upper limit.} 

We conclude that, while small arrays of a few low--timing-noise pulsars (represented here by the \citealt{s+15} configuration) provide the most constraining upper limits on the nanohertz GWB, they are suboptimal for detection in comparison to larger arrays such as NANOGrav, EPTA, full-PPTA, and the IPTA, which are expanded regularly with newly discovered pulsars.
Indeed, detection and upper limits have rather different demands, and the PPTA itself may employ a $\sim 20$-pulsar array for detection in the near future \citep{rhc+15}.
Encouragingly, we find that larger PTAs have $\sim 80\%$ probability of making a detection in the next ten years, even allowing for a reduced GWB signal due to significant binary stalling or environmental influences.

\section{Statistics} \label{sec:stats}
The details of the frequency-domain statistical framework used here can be found in \citet{rsg15}. In the frequentist context, a statistic $X$ is a single number that summarizes the (noisy) data with respect to the presence of a (GW) signal, so that, on average, a higher $A_\mathrm{true}$ yields a higher $X$, with fluctuations due to the range of possible noise realizations. Given a set of data, there is \emph{one} observed $\hat{X}$, and setting an upper limit $A_\mathrm{ul}^{95\%}$, amounts to stating that if $A_\mathrm{true}$ were equal to $A_\mathrm{ul}^{95\%}$, for 95\% of noise realizations the observed $X$ would have been higher than $\hat{X}$. 

Detection schemes based on statistics, on the other hand, proceed as follows: one considers separately the case where a signal is present in the data at a certain level, and the case where it is not, and computes the probability distribution of the statistic (over the ensemble of noise realizations) in both cases. The zero-signal distribution is used to set a \emph{detection threshold} as a function of a false-alarm probability; the signal distribution is used to compute the probability that for a certain $A_\mathrm{true}$ the statistic will exceed the threshold---that is, the detection probability (or efficiency). 

In the following we use cross-correlation statistics for both upper-limit and detection considerations, making use of the correlated influence of a GWB signal on pulsars which are widely separated on the sky.

{\it Upper limits.}-- For the discussion of upper limits in this work we will use the so-called $A$-statistic of \citet{rsg15} modified so that the standard deviation is measured in units of squared GWB amplitude and that the value of the statistic is equal to the injected GWB amplitude on average, as is done in \cite{ccs+15}. Assuming that the distribution of this statistic is Gaussian (a reasonable approximation at low signal-to-noise ratios \citep{ccs+15}) we obtain
\be
p(A^{2}|A_{\rm true}) = \frac{1}{\sqrt{2\pi\sigma_{0}^{2}}}\exp\left(-\frac{(A^{2}-A_{\rm true}^{2})^{2}}{2\sigma_{0}^{2}}\right)
\ee
with
\be
\sigma_{0} = A_{\rm true}^{-2}\left(  2\sum_{k}\sum_{ij}\Gamma_{ij}^{2}\frac{S_{h0}^{2}}{P_{i}P_{j}} \right)^{1/2},
\label{eq:asig}
\ee
where $\sum_{k}$ denotes a sum over frequencies $f_{k}$, $\sum_{ij}$ denotes a sum over pulsar pairs, $\Gamma_{ij}$ is the overlap reduction function (i.e. this is the \citet{hd83} curve for an isotropic GWB) for pulsars $i$ and $j$,  $S_{h0}\equiv S_{h0}(f_{k})$ is the modeled cross-power spectral density, and $P_{i}\equiv P_{i}(f_{k})$ is the intrinsic noise power spectral density. From this Gaussian probability distribution, it can be shown that
\be
A_{\rm ul}^{2} = \hat{A}^{2} + \sqrt{2}\sigma_{0}{\rm erfc}^{-1}(2(1-C)),
\ee
where $\hat{A}$ is the measured value of the GWB amplitude and $C$ is our upper limit confidence (e.g., 95\%). Thus, the distribution $p(A_{\rm ul}^{2}|A_{\rm true})$ is a Gaussian distribution with standard deviation $\sigma_{0}$ and mean $A_{\rm true}^{2}+\sqrt{2}\sigma_{0}{\rm erfc}^{-1}(2(1-C))$. Note, to compute $p(A_{\rm ul}|A_{\rm true})$ we perform a coordinate transformation to obtain $p(A_{\rm ul}|A_{\rm true})$ = $2A_{\rm ul}p(A_{\rm ul}^{2}|A_{\rm true})$.

\begin{figure}
\includegraphics[angle=0, width=0.5\textwidth]{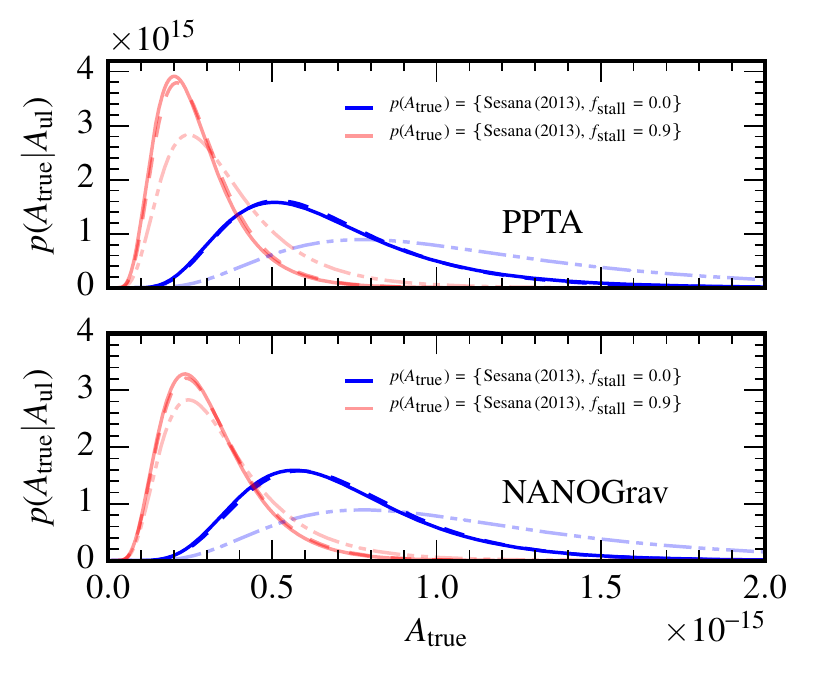}
\vspace{-18pt}
\caption{The normalized probability distribution for the SMBHB GW amplitude $A_\mathrm{true}$, given upper limits from the PPTA \citep{s+15} and a large PTA which regularly adds pulsars (specifically NANOGrav \citep{arz+15}). The darker blue curves assume an amplitude prior based on \citet{s13}; the faded red curves modify it by assuming 90\% binary stalling, reducing the amplitude $\sqrt{10}$-fold. The dashed curves reflect turnover spectra due to binary stellar hardening. The dash-dotted curves correspond to the astrophysically-motivated priors on the amplitude for no stalling (blue) and $90\%$ stalling (red).}
\label{fig:pAtruegivenAul}
\end{figure}

{\it Detection.}-- In the previous section we used the $A$-statistic for upper limits in order to compare our results with upper limits computed using the statistic presented in \cite{ccs+15}. For assessing our detection probability, however, we will make use of the $B$-statistic of \cite{rsg15}. In this case, we wish to determine the detection probability
\be
{\rm DP}=\frac{1}{2}{\rm erfc}\left[ \frac{\sqrt{2}\sigma_{0}{\rm erfc}^{-1}(2\alpha_{0})-\mu_{1}}{\sqrt{2}\sigma_{1}} \right],
\label{eq:dp}
\ee
where $\alpha_{0}$ is the false alarm probability (FAP); $\mu_{1}$, $\sigma_{1}$, and $\sigma_{0}$ (distinct from previous) are the mean in the presence of a signal, the standard deviation in the presence (subscript 1) and the standard deviation in the absence (subscript 0) of a signal (see Eqs.\@~A16--A18 of \citealt{rsg15} for more details). Both $\mu_1$ and $\sigma_1$ grow with increasing signal amplitude. Here we choose a FAP of $0.13\%$ to match the 3-sigma detection threshold of \citet{sejr13}.\footnote{This choice is clear if one notices that the 3-sigma range is a \emph{two}-sided confidence limit, thus 0.27\% of the probability density function is outside of this range. To convert to a standard FAP, we need the one-sided limit which is simply 0.27/2 $\sim$ 0.13.}

{\it Implementation.}-- In the work of \cite{rsg15}, it is argued that timing model subtraction (namely subtraction of the quadratic term of the timing model;  due to pulsar spin down) can be emulated by not including the lowest frequency in the sum over $k$ frequencies. However, we have found that these analytic results agree much better with simulations including timing model subtraction if all frequencies are included in the sum. Furthermore, to emulate data sets with different time spans we only include frequencies in the sum that are \emph{greater} than $1/T_{\rm min}$, where $T_{\rm min}$ is the shortest time span for the given pulsar pair.

In this work we always include the GWB power spectral density in the \emph{intrinsic} pulsar noise $P_{i}$. This is meant to mimic a real optimal statistic analysis where the intrinsic noise would be based on single-pulsar-analyses in which there is no way to distinguish the intrinsic noise from GWB power. This will make our analysis conservative in the sense that we are overestimating the intrinsic noise.

\section{Results \& discussion}

In the following, we characterize detection prospects in terms of small and large PTAs, and specifically consider the following five configurations:
\begin{enumerate}
\item PPTA4, consisting of the four pulsars described in \citet{s+15} with associated measurement (white) and timing (red) noise \citep{s+15,rhc+15}.
\item NANOGrav$+$, consisting of the $37$ pulsars described in Table $3$ of \citet{arz+15b} with their associated noise properties, to be expanded in future observations by adding four new pulsars each year with $250$-ns TOA measurement noise, in line with current expectations.
\item EPTA$+$, consisting of the $42$ pulsars described in Table $3$ of \citet{cll+15} with their associated noise properties, and regularly adding pulsars as in $(2)$. 
Some of these additional pulsars are monitored with the Large European Array for Pulsars (LEAP; \citealt{kc13,bassa15}), which synthesizes a 194-m steerable dish from the five EPTA telescopes.
\item A conservative IPTA+, consisting of the $49$ pulsars of the first IPTA data release, described in \citet{vlh+15} with measurement and timing noise\footnote{For IPTA pulsars timing noise is taken to consist only of red spin noise, and not of any other system- or band-specific red noise. The latter components may be isolated with multi-system and multi-frequency observations, while the former is completely conflated with a GWB signal in the absence of cross-correlation measurements.} properties as in \citet{lsc+15}, and expanding by six new pulsars each year, again with $250$-ns TOA measurement noise.
\item A theorist's PTA (TPTA), consisting of the toy configuration of \citet{rsg15} which may be supported by an advanced radio telescope such as SKA$1$: $50$ pulsars with $100$-ns TOA measurement noise and no intrinsic timing noise.
\end{enumerate}
These configurations are summarized in Table~\ref{tab:array_properties}.\footnote{We rescale PPTA, NANOGrav, EPTA, and IPTA measurement noises (but not the timing noise) by a common factor for each PTA, so that the corresponding $A$-statistic upper limits match the results of \citet{s+15}, \citet{arz+15}, \citet{ltm+15}, and \citet{vlh+15}; this rescaling has little impact on detection probability in the future, since white measurement noise becomes subdominant to red timing noise at low frequencies.}

\begin{table*}
\caption{PTAs considered, as characterized in our analysis.}
\label{tab:array_properties}
\centering
\scriptsize
\begin{tabular}{l@{\;\;}l@{\;\;}l}
\hline
PTA & configuration & number of pulsars \\ 
\hline
\hline
PPTA$4$ & PPTA \citep{s+15} & 4 \\[2ex]
NANOGrav$+$ & NANOGrav & $37$ $+$ $4$ new/year with $250$-ns\\
& \citep{arz+15b} & TOA measurement precision\\[2ex]
EPTA$+$ & EPTA \citep{cll+15} & $42$ $+$ $4$ new/year with $250$-ns\\
& & TOA measurement precision\\[2ex]
IPTA$+$ & IPTA Data Release $1$ & $49$ $+$ $6$ new/year with $250$-ns\\
& \citep{vlh+15} & TOA measurement precision\\
& \citep{lsc+15} & \\[2ex]
Theorist's PTA & $100$-ns TOA measurement precision & $50$\\
\hline\hline
\end{tabular}
\end{table*}

We consider four combined estimates for the amplitude and spectral shape of the stochastic GWB from SMBHBs, ranging from more optimistic (detection wise) to more conservative.
For the amplitude, we adopt the observationally motivated log-normal distribution of \citet{s13} with mean $\log_{10}A_\mathrm{true}=-15$ and standard deviation of $0.22$; however, we also consider the case where $90\%$ of binaries require longer than $\sim 9 - 13.7$ Gyr before reaching the PTA band (i.e., the binaries stall, \citealt{bs15,mop14}), which corresponds to a mean $\log_{10}A_\mathrm{true}=-15.5$, with the same standard deviation. For each of these amplitude priors, we examine the purely GW-driven power-law spectrum of Eq.\ \eqref{eq:powerlaw}, as well as the case where the GW strain has a turnover at $f = 1/(11\,\mathrm{yr})$ caused by SMBHB interactions with stars in galactic nuclei, which accelerate binary inspiral and remove low-frequency power with respect to the pure power law (see, e.g., \citealt{scm15} and \citealt{arz+15}). The $1/(11\,\mathrm{yr})$ turnover frequency was chosen to lie just beyond the low-frequency reach of the current PPTA dataset, since the \citet{s+15} analysis still found $\sim 9\%$ consistency with the purely GW-driven model of \citet{s13} for $f\geq 1/(11\,\mathrm{yr})$. Similarly, the recent NANOGrav \citep{arz+15} analysis finds 20\% consistency using a 9-year dataset \citep{arz+15b}. Furthermore, our choice of a turnover at $1/(11\,\mathrm{yr})$ corresponds to reasonable assumptions about the density of stellar populations interacting with SMBHBs in galactic nuclei (see Figs. 10 and 11 of \citealt{arz+15}). 
At very low frequencies, the frequency scaling of the timing-residual spectrum becomes $S(f)\propto 1/f$ due to the environmental influences. 

In the top panel of {Fig.}~\ref{fig:pAtruegivenAul} we show the probability distribution $p(A_\mathrm{true}|A_\mathrm{ul})$ obtained by setting the 95\% $A$-statistic upper limit to the \citet{s+15} value, $1\times 10^{-15}$, and by assuming no-stalling (solid blue line) and 90\%-stalling (faded solid red line) amplitude priors, with pure power-law true spectra. The dashed lines correspond to turnover spectra.\footnote{When analyzing these latter cases, we use the ``wrong'' spectrum -- a pure power law -- for the $P_i$ in Eq.\ \eqref{eq:asig}, for consistency with existing analyses that assume GW-driven inspirals in deriving limits. As discussed in \citet{rsg15}, doing so has negligible effects on the distribution of the $A$-statistic.} For comparison, in the bottom panel of {Fig.}~\ref{fig:pAtruegivenAul} we show also $p(A_\mathrm{true}|A_\mathrm{ul})$ as computed for NANOGrav's optimal-statistic upper limit $1.3\times 10^{-15}$ \citep{arz+15}. The underlying astrophysically-motivated priors for no-stalling and $90\%$-stalling are shown in each panel as blue and red dash-dot lines, respectively.

\begin{figure*}
\centering 
\includegraphics[angle=0, width=\textwidth]{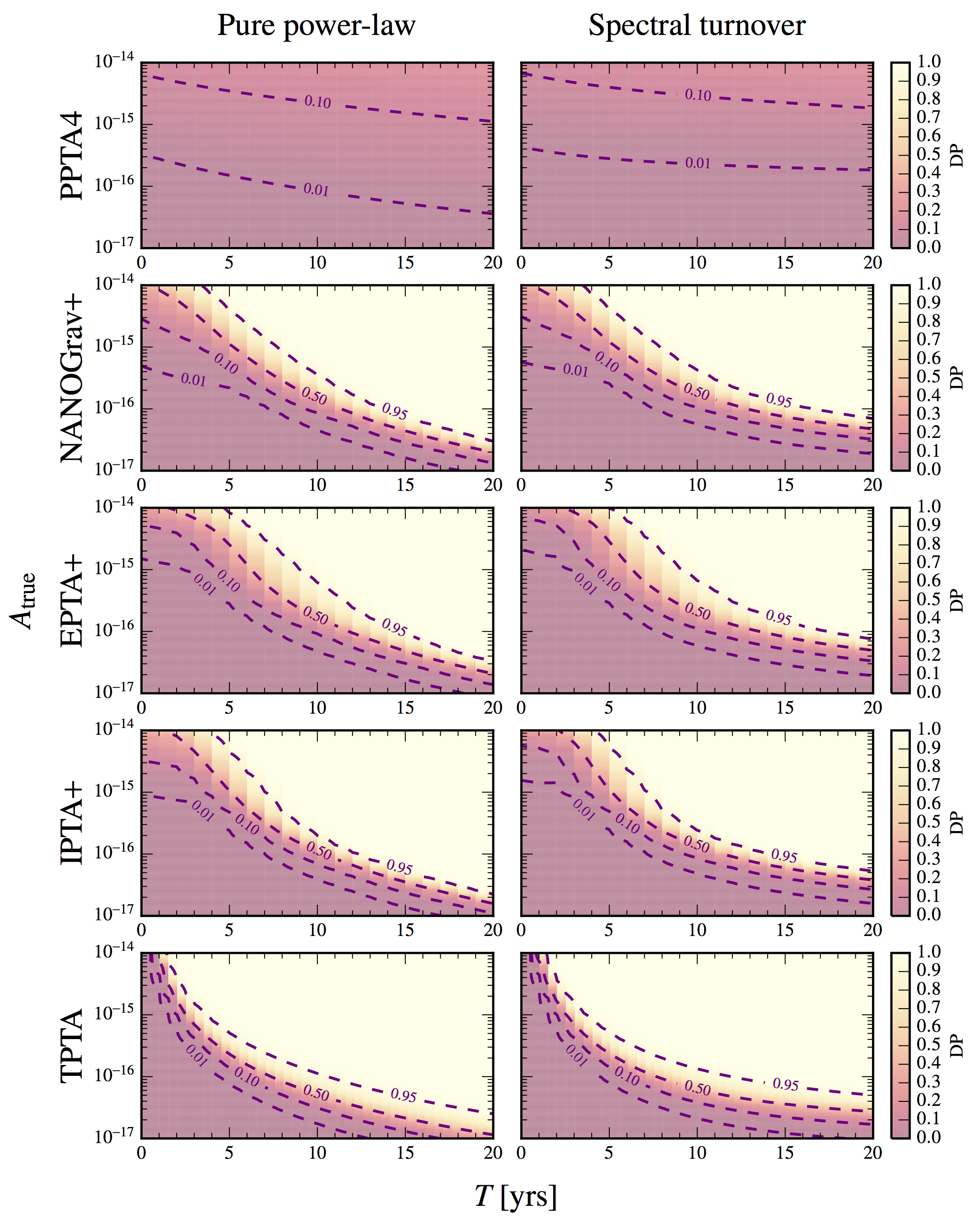}
\caption{Detection probabilities (DPs) for all arrays considered in this work, as a function of the true GW background amplitude $A_\mathrm{true}$ and of observation time $T$ beyond the existing dataset. The left panels were derived for pure power-law GW background spectra, and the right panels for turnover spectra with a knee at $f=1/(11 \,\mathrm{yrs})$ due to stellar hardening.
}
\label{fig:otherpta_dp}
\end{figure*}

Next, we compute the $B$-statistic GW detection probabilities, Eq.\ \eqref{eq:dp}, for each PTA, as a function of $A_\mathrm{true}$ and of the observation time $T$ beyond current datasets. The resulting DPs are shown in {Figs.}\ \ref{fig:otherpta_dp}: the left and right columns correspond to the pure power-law and turnover spectra, respectively.\footnote{In computing the $B$-statistic, we do assume that the analysis accounts correctly for the true spectral shape. Again, this choice has minimal effects.}

As mentioned above, for NANOGrav$+$/EPTA$+$ and IPTA+ we add new pulsars at rates of four and six per year, respectively. This annual expansion has a positive impact on the DP, since it provides more independent pulsar pairs at differing angular separations, which help discriminate GWs with quadrupolar spatial correlations (the \citealt{hd83} curve, or the more general anisotropic signatures discussed by \citealt{msmv13} and \citealt{grtm14}) from non--spatially-correlated red noise processes \citep{sejr13}. The unique GW correlation signature is the smoking gun for detection, and we must coordinate future efforts to maximize its measurement.

\begin{figure}
\includegraphics[angle=0, width=0.5\textwidth]{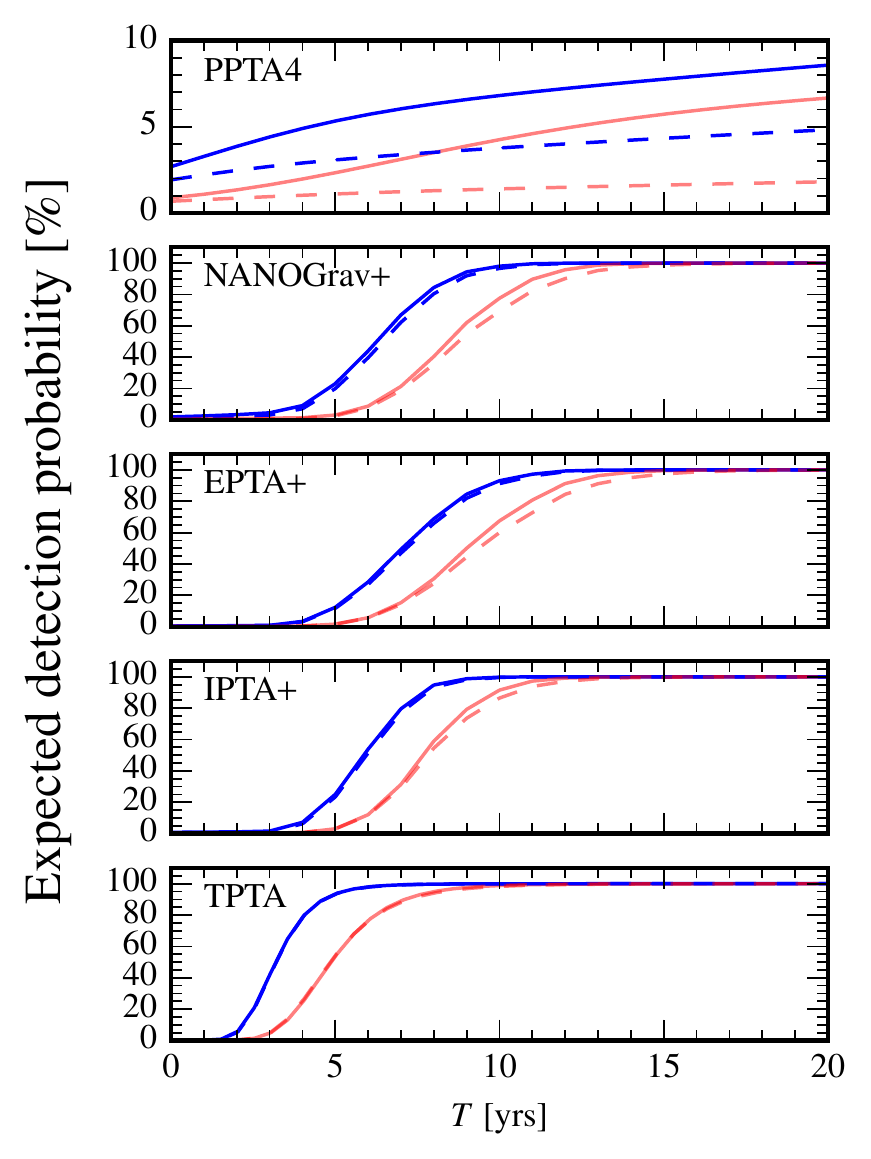}
\caption{Summary of results for the growth of GWB detection probability (DP) with further observation time in each array. Blue and red lines are for zero and $90\%$ binary stalling, respectively. Dashed lines correspond to the case where the true background spectrum has a turnover at $f=1/(11 \,\mathrm{yrs})$ due to binary stellar hardening.}
\label{fig:summary}
\end{figure}

We obtain the \emph{expected} detection probabilities $\langle \mathrm{DP} \rangle_{A_\mathrm{true}}$ as a function of time by integrating $\mathrm{DP}(A_\mathrm{true},T)$ against the $p(A_\mathrm{true}|A_\mathrm{ul})$ curves of {Fig.}\ \ref{fig:pAtruegivenAul} (top panel). The resulting $\langle \mathrm{DP} \rangle_{A_\mathrm{true}}$ are shown in {Figs.}\ \ref{fig:summary}. As in {Fig.}\ \ref{fig:pAtruegivenAul}, the darker blue and lighter red curves correspond to no-stalling and 90\%-stalling amplitude priors. For the PPTA$4$ array, expected detection probabilities remain below $10\%$ throughout the next $20$ years of observation, even in the most optimistic GW-signal scenario. This is unsurprising---a PTA consisting of a few exquisitely timed pulsars may provide very tight upper limits, but it will not usually yield convincing detection statistics, which require an array of many pulsars to map out the expected spatial correlation signature. 

By contrast, large pulsar arrays such as NANOGrav$+$, EPTA$+$, and IPTA$+$ provide high detection probabilities even with strong binary environmental couplings, since they allow the quadrupolar spatial correlation signature of a GWB to be mapped by many different pulsar pairings. We expect these results to be qualitatively the same for a full PPTA that regularly adds pulsars to the array. In NANOGrav$+$, EPTA$+$, and IPTA+, detection probabilities begin to grow rapidly after only five years of observation beyond current datasets. 
Binary stalling (the red curves) stunts this growth by three years at most. While the low-frequency turnover reduces detection probability, its effect is mitigated by the large number of pulsars and the annual catalog expansion.\footnote{\citet{s+15} advocate performing observations with higher cadence to improve sensitivity at $f\geq 0.2$ yr$^{-1}$, where the GWB would be less affected by environmental couplings. In fact, higher-cadence observations would improve sensitivity \emph{across all frequencies}, but not if they come at the cost of reducing the number of monitored pulsars because of limited observing-time allocations. Furthermore, GW searches are actually sensitive to the spectrum of timing residuals rather than of strain itself. The two are related by $S_h(f)\propto h_c(f)^2/f^3$, so even a turnover spectrum will leave a steep red-noise signature in the pulsar timing residuals.}

Finally, we see that the large number of well-timed pulsars in the TPTA builds highly convincing detection probabilities after only a few years of operation. By the time the pulsars have been observed long enough that the influence of the turnover at $f=1/(11\,\mathrm{yr})$ may be noticeable, the DP is already close to unity. The same is true were we to add timing noise at currently known levels to these TPTA pulsars--the DP will already be close to unity when low-frequency timing noise begins to have a significant influence. 

We stress that there are several important caveats to our analysis. We have focused on detecting the stochastic background rather than deterministic signals. Even for these, using data from more pulsars is desirable, in order to build evidence for a coherent signal. We have assumed for each pulsar a single value of measurement noise that does not improve over time (as occurs in reality when receivers and backends are upgraded), nor do we consider interstellar medium effects such as dispersion (see, e.g., \citealt{stinebring2013}). The influence of a GWB spectral turnover depends on its frequency, which is a function of the typical environmental properties of galactic nuclei (either directly or in how these properties drive SMBHB orbital eccentricity)---our choice corresponds to reasonable assumptions about the stellar mass density of SMBHB environments. The timelines for the growth of detection in each PTA are approximate, intended to emphasize the differences between PTAs suited to upper-limits versus detection, and to demonstrate the influence of various binary stalling and environmental scenarios on detection probabilities.

We conclude by emphasizing the different demands of placing stringent upper limits on the stochastic background versus actually detecting it. To wit: 
\begin{itemize}
\item Highly constraining, astrophysically significant upper limits are achievable with only a few exquisitely timed pulsars, but such a PTA is suboptimal for the detection of a stochastic GW background.
\item Timing many pulsars allows for the quadrupolar spatial correlation signature of the SMBHB background to be sampled at many different angular separations, enhancing prospects for detection.
\item Adding more pulsars regularly to PTAs will continually improve detection probability, in addition to the gains already made by timing existing pulsars for longer, and will help to mitigate the deleterious influences of binary stalling and environmental couplings.
\end{itemize}

\section{Acknowledgements}
It is our pleasure to thank Pablo Rosado, Alberto Sesana, Jonathan Gair, Lindley Lentati, Sarah Burke-Spolaor, Xavier Siemens, Maura McLaughlin, Joseph Romano, and Michael Kramer for very useful suggestions. We also thank the full NANOGrav collaboration for their comments and remarks. SRT was supported by an appointment to the NASA Postdoctoral Program at the Jet Propulsion Laboratory, administered by Oak Ridge Associated Universities through a contract with NASA. MV acknowledges support from the JPL RTD program. JAE and RvH acknowledge support by NASA through Einstein Fellowship grants PF4-150120 and PF3-140116, respectively. CMFM was supported by a Marie Curie International Outgoing Fellowship within the European Union Seventh Framework Programme. This work was supported in part by National Science Foundation Physics Frontier Center award no.\ 1430284, and by grant PHYS-1066293 and the hospitality of the Aspen Center for Physics. This research was performed at the Jet Propulsion Laboratory, under contract with the National Aeronautics and Space Administration. 



\bibliographystyle{apj}
\bibliography{bib}

\begin{thebibliography}{35}
\expandafter\ifx\csname natexlab\endcsname\relax\def\natexlab#1{#1}\fi

\bibitem[{{Arzoumanian} {et~al.}(2015{\natexlab{a}}){Arzoumanian}, {Brazier},
  {Burke-Spolaor}, {Chamberlin}, {Chatterjee}, {Christy}, {Cordes}, {Cornish},
  {Demorest}, {Deng}, {Dolch}, {Ellis}, {Ferdman}, {Fonseca}, {Garver-Daniels},
  {Jenet}, {Jones}, {Kaspi}, {Koop}, {Lam}, {Lazio}, {Levin}, {Lommen},
  {Lorimer}, {Luo}, {Lynch}, {Madison}, {McLaughlin}, {McWilliams},
  {Mingarelli}, {Nice}, {Palliyaguru}, {Pennucci}, {Ransom}, {Sampson},
  {Sanidas}, {Sesana}, {Siemens}, {Simon}, {Stairs}, {Stinebring}, {Stovall},
  {Swiggum}, {Taylor}, {Vallisneri}, {van Haasteren}, {Wang}, \&
  {Zhu}}]{arz+15}
{Arzoumanian}, Z., {et~al.} 2015{\natexlab{a}}, ArXiv e-prints, 1508.03024

\bibitem[{{Arzoumanian} {et~al.}(2015{\natexlab{b}}){Arzoumanian}, {Brazier},
  {Burke-Spolaor}, {Chamberlin}, {Chatterjee}, {Christy}, {Cordes}, {Cornish},
  {Crowter}, {Demorest}, {Dolch}, {Ellis}, {Ferdman}, {Fonseca},
  {Garver-Daniels}, {Gonzalez}, {Jenet}, {Jones}, {Jones}, {Kaspi}, {Koop},
  {Lam}, {Lazio}, {Levin}, {Lommen}, {Lorimer}, {Luo}, {Lynch}, {Madison},
  {McLaughlin}, {McWilliams}, {Nice}, {Palliyaguru}, {Pennucci}, {Ransom},
  {Siemens}, {Stairs}, {Stinebring}, {Stovall}, {Swiggum}, {Vallisneri}, {van
  Haasteren}, {Wang}, \& {Zhu}}]{arz+15b}
---. 2015{\natexlab{b}}, \apj, 813, 65

\bibitem[{{Bassa} {et~al.}(2016){Bassa}, {Janssen}, {Karuppusamy}, {Kramer},
  {Lee}, {Liu}, {McKee}, {Perrodin}, {Purver}, {Sanidas}, {Smits}, \&
  {Stappers}}]{bassa15}
{Bassa}, C.~G., {et~al.} 2016, \mnras, 456, 2196

\bibitem[{{Burke-Spolaor} \& {Simon}(2015)}]{bs15}
{Burke-Spolaor}, S., \& {Simon}, J. 2015, private communication

\bibitem[{{Caballero} {et~al.}(2015){Caballero}, {Lee}, {Lentati}, {Desvignes},
  {Champion}, {Verbiest}, {Janssen}, {Stappers}, {Kramer}, {Lazarus},
  {Possenti}, {Tiburzi}, {Perrodin}, {Os{\l}owski}, {Babak}, {Bassa}, {Brem},
  {Burgay}, {Cognard}, {Gair}, {Graikou}, {Guillemot}, {Hessels},
  {Karuppusamy}, {Lassus}, {Liu}, {McKee}, {Mingarelli}, {Petiteau}, {Purver},
  {Rosado}, {Sanidas}, {Sesana}, {Shaifullah}, {Smits}, {Taylor}, {Theureau},
  {van Haasteren}, \& {Vecchio}}]{cll+15}
{Caballero}, R.~N., {et~al.} 2015, ArXiv e-prints

\bibitem[{{Chamberlin} {et~al.}(2015){Chamberlin}, {Creighton}, {Siemens},
  {Demorest}, {Ellis}, {Price}, \& {Romano}}]{ccs+15}
{Chamberlin}, S.~J., {Creighton}, J.~D.~E., {Siemens}, X., {Demorest}, P.,
  {Ellis}, J., {Price}, L.~R., \& {Romano}, J.~D. 2015, \prd, 91, 044048

\bibitem[{{Gair} {et~al.}(2014){Gair}, {Romano}, {Taylor}, \&
  {Mingarelli}}]{grtm14}
{Gair}, J., {Romano}, J.~D., {Taylor}, S., \& {Mingarelli}, C.~M.~F. 2014,
  \prd, 90, 082001

\bibitem[{{Hellings} \& {Downs}(1983)}]{hd83}
{Hellings}, R.~W., \& {Downs}, G.~S. 1983, \apjl, 265, L39

\bibitem[{{Hobbs}(2013)}]{h13}
{Hobbs}, G. 2013, Classical and Quantum Gravity, 30, 224007

\bibitem[{Hobbs {et~al.}(2010)Hobbs, Archibald, Arzoumanian, Backer, Bailes,
  Bhat, Burgay, Burke-Spolaor, Champion, Cognard, Coles, Cordes, Demorest,
  Desvignes, Ferdman, Finn, Freire, Gonzalez, Hessels, Hotan, Janssen, Jenet,
  Jessner, Jordan, Kaspi, Kramer, Kondratiev, Lazio, Lazaridis, Lee, Levin,
  Lommen, Lorimer, Lynch, Lyne, Manchester, McLaughlin, Nice, Oslowski, Pilia,
  Possenti, Purver, Ransom, Reynolds, Sanidas, Sarkissian, Sesana, Shannon,
  Siemens, Stairs, Stappers, Stinebring, Theureau, van Haasteren, van Straten,
  Verbiest, Yardley, \& You}]{haa10}
Hobbs, G., {et~al.} 2010, Classical and Quantum Gravity, 27, 084013

\bibitem[{{Jaffe} \& {Backer}(2003)}]{jb03}
{Jaffe}, A.~H., \& {Backer}, D.~C. 2003, \apj, 583, 616

\bibitem[{{Janssen} {et~al.}(2015){Janssen}, {Hobbs}, {McLaughlin}, {Bassa},
  {Deller}, {Kramer}, {Lee}, {Mingarelli}, {Rosado}, {Sanidas}, {Sesana},
  {Shao}, {Stairs}, {Stappers}, \& {Verbiest}}]{2015aska.confE..37J}
{Janssen}, G., {et~al.} 2015, Advancing Astrophysics with the Square Kilometre
  Array (AASKA14), 37

\bibitem[{{Kramer} \& {Champion}(2013)}]{kc13}
{Kramer}, M., \& {Champion}, D.~J. 2013, Classical and Quantum Gravity, 30,
  224009

\bibitem[{{Lentati} {et~al.}(2013){Lentati}, {Alexander}, {Hobson}, {Taylor},
  {Gair}, {Balan}, \& {van Haasteren}}]{l+13}
{Lentati}, L., {Alexander}, P., {Hobson}, M.~P., {Taylor}, S., {Gair}, J.,
  {Balan}, S.~T., \& {van Haasteren}, R. 2013, \prd, 87, 104021

\bibitem[{{Lentati} {et~al.}(2015){Lentati}, {Taylor}, {Mingarelli}, {Sesana},
  {Sanidas}, {Vecchio}, {Caballero}, {Lee}, {van Haasteren}, {Babak}, {Bassa},
  {Brem}, {Burgay}, {Champion}, {Cognard}, {Desvignes}, {Gair}, {Guillemot},
  {Hessels}, {Janssen}, {Karuppusamy}, {Kramer}, {Lassus}, {Lazarus}, {Liu},
  {Os{\l}owski}, {Perrodin}, {Petiteau}, {Possenti}, {Purver}, {Rosado},
  {Smits}, {Stappers}, {Theureau}, {Tiburzi}, \& {Verbiest}}]{ltm+15}
{Lentati}, L., {et~al.} 2015, \mnras, 453, 2576

\bibitem[{Lentati {et~al.}(2016)Lentati, Shannon, Coles, Verbiest, van
  Haasteren, Ellis, Caballero, Manchester, Arzoumanian, Babak, Bassa, Bhat,
  Brem, Burgay, Burke-Spolaor, Champion, Chatterjee, Cognard, Cordes, Dai,
  Demorest, Desvignes, Dolch, Ferdman, Fonseca, Gair, Gonzalez, Graikou,
  Guillemot, Hessels, Hobbs, Janssen, Jones, Karuppusamy, Keith, Kerr, Kramer,
  Lam, Lasky, Lassus, Lazarus, Lazio, Lee, Levin, Liu, Lynch, Madison, McKee,
  McLaughlin, McWilliams, Mingarelli, Nice, Osłowski, Pennucci, Perera,
  Perrodin, Petiteau, Possenti, Ransom, Reardon, Rosado, Sanidas, Sesana,
  Shaifullah, Siemens, Smits, Stairs, Stappers, Stinebring, Stovall, Swiggum,
  Taylor, Theureau, Tiburzi, Toomey, Vallisneri, van Straten, Vecchio, Wang,
  Wang, You, Zhu, \& Zhu}]{lsc+15}
Lentati, L., {et~al.} 2016, \mnras

\bibitem[{{McLaughlin}(2013)}]{ml13}
{McLaughlin}, M.~A. 2013, Classical and Quantum Gravity, 30, 224008

\bibitem[{{McWilliams} {et~al.}(2014){McWilliams}, {Ostriker}, \&
  {Pretorius}}]{mop14}
{McWilliams}, S.~T., {Ostriker}, J.~P., \& {Pretorius}, F. 2014, \apj, 789, 156

\bibitem[{{Mingarelli} {et~al.}(2013){Mingarelli}, {Sidery}, {Mandel}, \&
  {Vecchio}}]{msmv13}
{Mingarelli}, C.~M.~F., {Sidery}, T., {Mandel}, I., \& {Vecchio}, A. 2013,
  Physical Review D, 88, 062005

\bibitem[{{Rajagopal} \& {Romani}(1995)}]{rm95}
{Rajagopal}, M., \& {Romani}, R.~W. 1995, \apj, 446, 543

\bibitem[{{Ravi} {et~al.}(2014){Ravi}, {Wyithe}, {Shannon}, {Hobbs}, \&
  {Manchester}}]{rws+14}
{Ravi}, V., {Wyithe}, J.~S.~B., {Shannon}, R.~M., {Hobbs}, G., \& {Manchester},
  R.~N. 2014, \mnras, 442, 56

\bibitem[{{Reardon} {et~al.}(2016){Reardon}, {Hobbs}, {Coles}, {Levin},
  {Keith}, {Bailes}, {Bhat}, {Burke-Spolaor}, {Dai}, {Kerr}, {Lasky},
  {Manchester}, {Os{\l}owski}, {Ravi}, {Shannon}, {van Straten}, {Toomey},
  {Wang}, {Wen}, {You}, \& {Zhu}}]{rhc+15}
{Reardon}, D.~J., {et~al.} 2016, \mnras, 455, 1751

\bibitem[{{Rosado} {et~al.}(2015){Rosado}, {Sesana}, \& {Gair}}]{rsg15}
{Rosado}, P.~A., {Sesana}, A., \& {Gair}, J. 2015, \mnras, 451, 2417

\bibitem[{{Sampson} {et~al.}(2015){Sampson}, {Cornish}, \&
  {McWilliams}}]{scm15}
{Sampson}, L., {Cornish}, N.~J., \& {McWilliams}, S.~T. 2015, \prd, 91, 084055

\bibitem[{{Sesana}(2013)}]{s13}
{Sesana}, A. 2013, \mnras, 433, L1

\bibitem[{{Sesana} {et~al.}(2004){Sesana}, {Haardt}, {Madau}, \&
  {Volonteri}}]{shm+04}
{Sesana}, A., {Haardt}, F., {Madau}, P., \& {Volonteri}, M. 2004, \apj, 611,
  623

\bibitem[{{Sesana} {et~al.}(2008){Sesana}, {Vecchio}, \& {Colacino}}]{svc08}
{Sesana}, A., {Vecchio}, A., \& {Colacino}, C.~N. 2008, \mnras, 390, 192

\bibitem[{{Shannon} {et~al.}(2015){Shannon}, {Ravi}, {Lentati}, {Lasky},
  {Hobbs}, {Kerr}, {Manchester}, {Coles}, {Levin}, {Bailes}, {Bhat},
  {Burke-Spolaor}, {Dai}, {Keith}, {Oslowski}, {Reardon}, {van Straten},
  {Toomey}, {Wang}, {Wen}, {Wyithe}, \& {Zhu}}]{s+15}
{Shannon}, R.~M., {et~al.} 2015, Science, 349, 1522

\bibitem[{{Siemens} {et~al.}(2013){Siemens}, {Ellis}, {Jenet}, \&
  {Romano}}]{sejr13}
{Siemens}, X., {Ellis}, J., {Jenet}, F., \& {Romano}, J.~D. 2013, Classical and
  Quantum Gravity, 30, 224015

\bibitem[{{Stinebring}(2013)}]{stinebring2013}
{Stinebring}, D. 2013, Classical and Quantum Gravity, 30, 224006

\bibitem[{{Vallisneri}(2012)}]{2012PhRvD..86h2001V}
{Vallisneri}, M. 2012, \prd, 86, 082001

\bibitem[{van Haasteren \& Levin(2013)}]{vl13}
van Haasteren, R., \& Levin, Y. 2013, Monthly Notices of the Royal Astronomical
  Society, 428, 1147

\bibitem[{{van Haasteren} {et~al.}(2009){van Haasteren}, {Levin}, {McDonald},
  \& {Lu}}]{vlml09}
{van Haasteren}, R., {Levin}, Y., {McDonald}, P., \& {Lu}, T. 2009, \mnras,
  395, 1005

\bibitem[{{Verbiest} {et~al.}(2016){Verbiest}, {Lentati}, {Hobbs}, {van
  Haasteren}, {Demorest}, {Janssen}, {Wang}, {Desvignes}, {Caballero}, {Keith},
  {Champion}, {Arzoumanian}, {Babak}, {Bassa}, {Bhat}, {Brazier}, {Brem},
  {Burgay}, {Burke-Spolaor}, {Chamberlin}, {Chatterjee}, {Christy}, {Cognard},
  {Cordes}, {Dai}, {Dolch}, {Ellis}, {Ferdman}, {Fonseca}, {Gair},
  {Garver-Daniels}, {Gentile}, {Gonzalez}, {Graikou}, {Guillemot}, {Hessels},
  {Jones}, {Karuppusamy}, {Kerr}, {Kramer}, {Lam}, {Lasky}, {Lassus},
  {Lazarus}, {Lazio}, {Lee}, {Levin}, {Liu}, {Lynch}, {Lyne}, {Mckee},
  {McLaughlin}, {McWilliams}, {Madison}, {Manchester}, {Mingarelli}, {Nice},
  {Os{\l}owski}, {Palliyaguru}, {Pennucci}, {Perera}, {Perrodin}, {Possenti},
  {Petiteau}, {Ransom}, {Reardon}, {Rosado}, {Sanidas}, {Sesana}, {Shaifullah},
  {Shannon}, {Siemens}, {Simon}, {Smits}, {Spiewak}, {Stairs}, {Stappers},
  {Stinebring}, {Stovall}, {Swiggum}, {Taylor}, {Theureau}, {Tiburzi},
  {Toomey}, {Vallisneri}, {van Straten}, {Vecchio}, {Wang}, {Wen}, {You},
  {Zhu}, \& {Zhu}}]{vlh+15}
{Verbiest}, J.~P.~W., {et~al.} 2016, \mnras

\bibitem[{{Wyithe} \& {Loeb}(2003)}]{wl03}
{Wyithe}, J.~S.~B., \& {Loeb}, A. 2003, \apj, 590, 691

\end{thebibliography}

 \end{document}